\documentclass[aps,prl,reprint,groupedaddress]{revtex4-2}

\usepackage{amsmath}
\usepackage{lipsum}
\usepackage{graphicx}
\graphicspath{{images/}}
\usepackage{dcolumn}
\usepackage{bm}

\usepackage[utf8]{inputenc}
\usepackage[ngerman,german,english]{babel}

\usepackage{amssymb}
\usepackage{soul}
\usepackage[dvipsnames]{xcolor}
\usepackage[normalem]{ulem}

\begin{document}

\title{Large speed-up of quantum emitter detection via quantum interference}
\author{Warwick P. Bowen}

\affiliation{%
 Australian Research Council Centre of Excellence in Quantum Biotechnology, School of Mathematics and Physics, University of Queensland, St Lucia, Queensland 4072, Australia.
}%

\date{\today}

\begin{abstract}
Quantum emitters are a key resource in quantum technologies, microscopy, and other applications. The ability to rapidly detect them is useful both for quality control in engineered emitter arrays and for high-contrast imaging of naturally occurring emitters. Using full photon-counting statistics and optimal Bayesian hypothesis testing, we show that extended Hong-Ou-Mandel interference between quantum emission and a coherent field enables orders-of-magnitude speed-ups in emitter detection under realistic noise and loss. Strikingly, the performance advantage improves as loss and background noise increase, and persists for incoherent emission. Taken together with prior demonstrations of extended Hong-Ou-Mandel interference, this suggest that substantial performance gains are achievable with current technology under realistic, non-ideal conditions. This offers a new approach to fast, low-intensity imaging and for emitter characterization in large-scale quantum systems. Fundamentally, the discovery that quantum interference and measurements, used together, are more robust to both loss and noise than standard measurement techniques opens the possibility of broad applications across quantum metrology.
\end{abstract}
\maketitle

Quantum emitters are central to applications ranging from emerging quantum technologies~\cite{aharonovich2016solid,defienne2024advances} to existing precision microscopes~\cite{vicidomini2018sted,li2020adaptive} and DNA sequencing~\cite{fodor1997dna}. Methods to rapidly detect them are important. For instance, scalable photonic quantum computers~\cite{psiquantum2025manufacturable} and repeaters~\cite{azuma2023quantum} require accurate characterization of large, high-fidelity arrays of single-photon sources; while, localisation of emitters with high speed and contrast is fundamental to many forms of microscopy~\cite{deschout2014precisely,rust2006sub,vicidomini2018sted}.

The most direct detection method is to excite the emitter and measure its spontaneous emission. However, due to background noise and detection losses, this can require many repetitions. It has been shown that quantum emission can exhibit {\it extended Hong-Ou-Mandel (HOM) interference} with coherent light~\cite{PhysRevA.86.063828, PhysRevA.105.013712, HOModd, rarity2005non,lyons2023fluorescence}. Simultaneously, photon-number-resolving detectors capable of measuring this non-classical interference are advancing rapidly~\cite{eaton2023resolution}, with integrated arrays now approaching thousands of pixels~\cite{cheng2023100, madonini2021single}. Motivated by these developments, we ask: can non-classical interference enable rapid detection of quantum emitters?

We find that the answer is yes. We derive the photon-number statistics for extended HOM interference~\cite{hong1987measurement,PhysRevA.105.013712}, including realistic imperfections. Via optimal Bayesian analysis, we show that the photon correlations this generates yield orders-of-magnitude improvements in detection speed over direct measurement. Unusually for quantum protocols, the advantage improves with increasing background noise and, when the emitter produces a coherent superposition of zero and one photon~\cite{loredo2019generation, maillette2024quantum}, with increasing loss. This is fundamentally interesting, showing that photon number resolving detectors can enable quantum measurements that are more robust than traditional detection methods, a result that may be of broad applicability. Also of fundamental interest, our results show that the existence of coherence makes it is easier to detect a superposition of zero and one photon than a (brighter) pure single photon state.

\begin{figure}[ht!]
	\includegraphics[width=7.5cm]{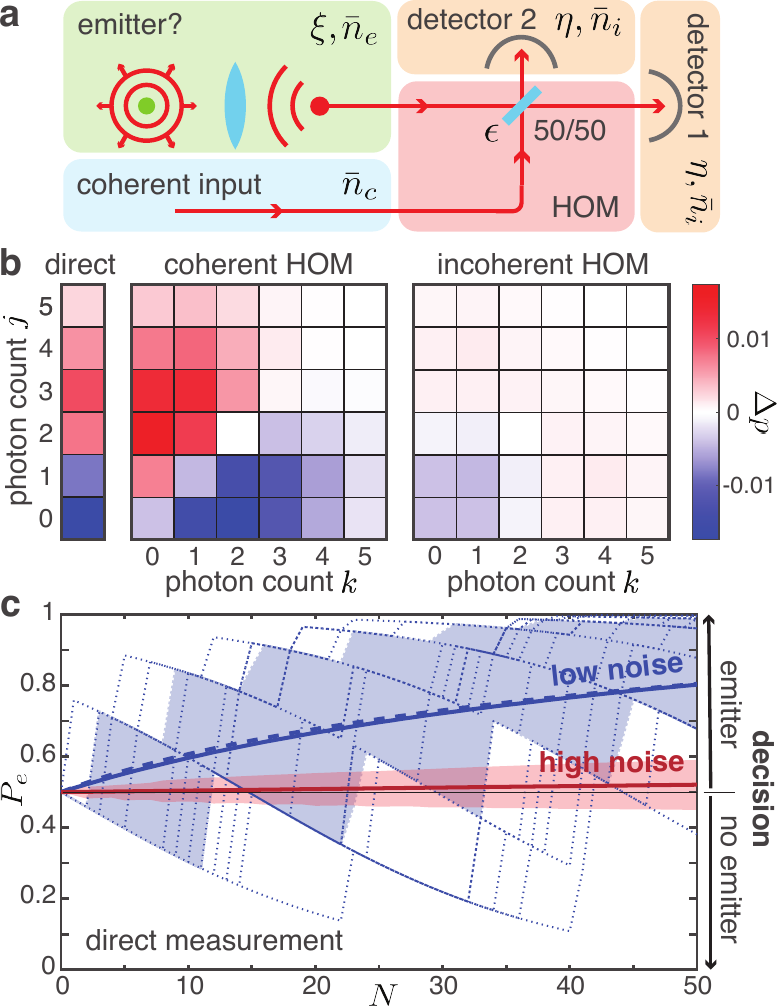}
	\caption{{\bf a.} Extended HOM-based emitter detection scheme. {\bf b.}  Photon count probability differences $\Delta p$ with and without an emitter. {\it Left:} $p_j(\xi)-p_j(0)$, with $p_j(\xi)$ the probability of observing $j$ photons via direct measurement. {\it Middle:} $p_{jk}(\xi)-p_{jk}(0)$ for optimal coherent HOM ($\cos \theta=1$). {\it Right:} $p_{jk}(\xi)-p_{jk}(0)$ for incoherent HOM ($\cos \theta=0$). $\bar{n}_e=\bar{n}_i=0.8$, 
	$\bar{n}_c=1$.  {\bf c.} Trajectories of $P_e$ for direct measurement. Blue: $\bar{n}_e=\bar{n}_i=0.02$; Red: $\bar{n}_e=\bar{n}_i=1$.
	Dotted lines: example trajectories. Solid lines: means of 100,000 trajectories. Dashed lines: means calculated using log-normal approximation. Shading:  $P_e$ range within lower and upper quartiles. {\bf b}\&{\bf c}: $\eta=0.8$.
	\label{fig:diagram} }
\end{figure}

Our proposed scheme for quantum interference-enhanced quantum emitter detection is experimentally feasible. We find that substantial speed-ups persist for incoherent emission and using simple photon-counting detectors, supporting practical implementation with existing technology. The scheme is robust to both inefficiencies and background noise and requires only that the spatiotemporal properties of the coherent field are well matched to those of the fluorescent emission, as demonstrated in Ref.~\cite{lyons2023fluorescence}. Moreover, HOM interference has already been applied in spectroscopy of organic molecules~\cite{eshun2021investigations}, phase-imaging microscopy~\cite{ndagano2022quantum}, and fluorescence lifetime microscopy~\cite{li1997femtosecond,lyons2023fluorescence}, and has been implemented in camera-based imaging schemes~\cite{jachura2015shot, ndagano2022quantum}. 

Together, our results open a pathway to enhance imaging techniques such as fluorescence and nonlinear microscopy
-- where fast, low-intensity measurements are essential~\cite{li2020adaptive,casacio2021quantum,lyons2023fluorescence}, and to enable rapid characterization of emitter arrays for scalable quantum technologies~\cite{psiquantum2025manufacturable,kielpinski2002architecture}.

{\bf Photon statistics of extended HOM:} Extended HOM interference 
causes anti-bunching of photon counts between two interferometer outputs~\cite{hong1987measurement,PhysRevA.105.013712}. We wish to exploit this to distinguish genuine quantum emission from background photon and detector counts. 
We consider the scheme in Fig.~\ref{fig:diagram}{\bf a}. The quantum emission is modelled as the coherent superposition $| e \rangle  =  \sqrt{1-\xi} |0 \rangle + \sqrt{\xi} |1 \rangle$, where
$\xi$ is the probability of photon emission. It
interferes, with imperfect mode overlap $\epsilon$,
with a coherent field of average photon occupancy $\bar n_c$. The interferometer outputs are detected on a pair of photon-number resolving detectors, each with efficiency $\eta$ and Poissionian electronic noise that introduces spurious counts with mean $\bar n_i$. Background noise in the vicinity of the emitter, for example from autofluorescence~\cite{waters2009accuracy} or pump leakage~\cite{piekarek2017high}, is modelled as populating an orthogonal spatiotemporal mode to the quantum emission and propagating independently through the apparatus to the detectors. This is also assumed to be Poissionian, with mean photon number $\bar n_e$ at the emitter location. 

Propagating the HOM input fields and vacuum states from two input loss channels through the apparatus 
 results in the output state (Sections 1\&3, Supplementary Material~\cite{SI}):
\begin{widetext}
\begin{equation}
\left | \psi_{\rm out} \right \rangle = e^{-\epsilon \bar{n}_c/2} \sum_{j,k,p,q=0}^\infty \frac{\alpha_{d}^{j+k} \alpha_{l}^{p+q}}{\sqrt{j! \, k! \, p! \, q!}}  \left [ \sqrt{1-\xi} + \sqrt{\frac{\xi}{\epsilon}}\left ( \frac{k-j + q-p}{\alpha_c}\right )\right ] | j \, k \,  p \, q \rangle,
\label{wavefunct}
\end{equation}
\end{widetext}
where $j$ and $k$ label the modes reaching each detector, $p$ and $q$ label inaccessible output loss channels, 
 $\alpha_c = \sqrt{\bar{n}_c} e^{i \theta}$ is the coherent amplitude of the 
 input coherent field, with $\theta$ its phase,
and $\alpha_d = \sqrt{\eta \epsilon/2} \, \alpha_c$ and $\alpha_l =  \sqrt{(1-\eta) \epsilon/2} \, \alpha_c$ are the input coherent amplitudes 
reaching the detector and loss outputs, respectively. 
An interesting observation from Eq.~(\ref{wavefunct}) is that, in limit of perfect single photon emission and detection efficiency ($\xi=\eta=1$), the sum over loss channels drops out ($\alpha_l^{p+q}/\sqrt{p! q!} \rightarrow \delta_{p,0} \delta_{q,0} $), resulting in zero probability amplitude when $j=k$. I.e., even with one coherent input, if the other is a single photon the interferometer outputs cannot contain equal numbers of photons. This is a manifestation of the {\it extended HOM effect}~\cite{PhysRevA.86.063828, PhysRevA.105.013712, HOModd}.

Tracing over the output loss channels and convolving with the Poissonian background noise we obtain the probability distribution of detected counts~\cite{SI}:
\begin{widetext}
\begin{equation}
{p}_{jk}(\xi) = \frac{e^{-\bar{n}}}{j! \, k!} \left (\frac{\bar{n}}{2} \right )^{ \! j+k} \left \{ 1 - \eta \xi \left [1 -  \frac{\bar{n}_n}{\bar{n}} \! \left (\frac{j+k}{\bar{n}}  \right )  - \eta \epsilon \, \bar{n}_c \! \left ( \frac{j-k}{\bar n} \right )^{\! \! 2}  \! + 2 \cos \theta \sqrt{\left ( \frac{1-\xi}{\xi} \right ) \epsilon \, \bar{n}_c} \left ( \frac{j-k}{\bar{n}}\right )  \right ]  \right \}, \label{P_coh}
\end{equation}
\end{widetext}
where $\bar{n} =  \eta \epsilon \bar{n}_c + \bar{n}_n$ is the total mean photon number detected across both detectors in the absence of an emitter, and
$\bar{n}_n = \eta(1-\epsilon) \bar{n}_c + \eta \bar{n}_e + 2 \bar{n}_i$ the total mean noise photon number, with the $\eta(1-\epsilon) \bar{n}_c$ term accounting for shot noise introduced by imperfect overlap.
This result goes beyond probability distributions previous derived for extended HOM interference~\cite{PhysRevA.86.063828, PhysRevA.105.013712, HOModd}. First, it provides the photon statistics for quantum emission into a coherent superposition of zero and one photon. 
Second, while imperfect detection efficiency has  recently been considered~\cite{HOModd}, only the diagonal terms ${p}_{jj}$ were calculated and these 
were expressed in terms of an infinite sum. Finally, to our knowledge previous work has not included background and detector noise sources. 

Our approach to improve emitter detection harnesses the quantum interference evident in the last two terms in square brackets in Eq.~(\ref{P_coh}), one phase sensitive and the other phase insensitive. Their $(j-k)$ dependences cause HOM-like anti-bunching of output photon counts.
For {\it coherent} extended HOM, where the emitter produces a superposition of zero and one photon, we find that the phase sensitive term should be maximised, setting  $\cos \theta=1$ henceforth. For {\it incoherent} extended HOM, where the emitter produces a mixed state 
-- such as fluorescent emission --  the probability distribution is obtained by averaging over all possible phases, or equivalently setting  $\cos \theta=0$. The phase sensitive term then disappears, but anti-bunching remains from the phase-insensitive term.  Interestingly, the phase sensitive 
 term is linear in $\eta$ while the phase-insensitive term is quadratic. This suggests that coherent HOM interference may be more robust to inefficiencies. The phase sensitive term also disappears for pure single photon emission ($\xi=1$), suggesting that superposition states may be easier and more robust to detect than single photons.

To decide whether an emitter is present, the outcomes of a set of measurements are compared
to the expected probability distributions with emitter present ($\xi \neq 0$) or absent ($\xi = 0$). The larger the difference between the distributions,
the more confidence we can expect in our decision.  Fig.~\ref{fig:diagram}{\bf b} plots the difference 
for each of the three relevant protocols: direct measurement on a single photon number resolving detector (see Section 1, Supplementary Material~\cite{SI}), coherent HOM, and incoherent HOM. Here and henceforth, we set $\xi=0.1$, $\epsilon=0.9$ and $\bar{n}_i=\bar{n}_e$ to constrain the parameter space. For direct measurement, as expected, the presence of an emitter increases the chance of observing a larger number of photons. By contrast, coherent HOM rebalances photons between interferometer arms, resulting in a substantially higher difference in probabilities.
 With two detectors, it also has many more possible measurement outcomes with high probability differences.
  Thus, we can expect better prediction power. Incoherent HOM washes out the rebalancing of photons,
  as might be anticipated. This greatly reduces the probability difference, with a maximum far lower than  direct measurement. This is offset by the larger number of possible measurement outcomes, making it unclear whether incoherent HOM will improve or degrade prediction power.

{\bf Bayesian hypothesis testing:} To optimally decide whether an emitter is present and quantify our confidence in this decision, we employ Bayesian hypothesis testing. Each run of an experiment provides two numbers: the measured photon counts $j$ and $k$ on the two detectors. The key statistical parameter obtained is the {\it likelihood ratio}~\cite{good1989weight}, defined as the ratio of conditional probabilities of the measurement outcome in the presence and absence of the emitter: $\lambda_{jk}  = {{p}_{jk}(\xi) }/{{p}_{jk}(0)}$.
A likelihood ratio above (below) 1 indicates that the emitter is more likely to be present (absent), with a larger (smaller) value indicating higher confidence. After a series of $N$ such independent measurements, Bayesian inference gives the conditional probability $P_e$ that the emitter is present~\cite{SI}. 
Assuming no prior information about its presence
\begin{equation}
P_e  =   \frac{1}{1+  \Lambda}, \label{eq_Pe}
\end{equation}
 where $\Lambda = \prod_{m=1}^N \lambda_{jk}^{(m)}$ is the {\it overall likelihood ratio} and $m$ is the measurement number.
We decide that the emitter is present if $P_e>0.5$ and absent if $P_e<0.5$.

One way to determine the confidence in this decision for a given protocol is to repeatedly simulate measurement outcomes and calculate $P_e$ for each. The confidence equals the fraction of simulations that yield the correct decision. As an illustration, we do this for direct measurement in the presence of an emitter~\cite{SI}.
Here, the total background noise is $\bar{n}_n^{\text{direct}}  =  \eta \bar{n}_e + \bar{n}_i$, lower than for extended HOM since only one detector is needed and, with no interference, there is no mode-mismatch noise. Fig.~\ref{fig:diagram}{\bf c} shows the resulting {\it measurement trajectories} of $P_e$ for two 
different background noise levels.

In the first simulation (blue traces, Fig.~\ref{fig:diagram}{\bf c}) the mean background noise is low enough ($\bar{n}_e=\bar{n}_i=0.02$)
that emitter photons dominate. The dotted lines are individual measurement trajectories, evidencing discrete jumps 
each time a photon is detected. The solid blue curve is the average over $100,000$ trajectories, with shaded region showing the upper and lower quartiles. Our confidence that the emitter exists increases as more measurements are made. After 50 measurements, 84\% of the trajectories give the correct decision ($P_e>0.5$), corresponding to a confidence $c_e=0.84$.
The second simulation (red traces, Fig.~\ref{fig:diagram}{\bf c}) has increased background noise ($\bar{n}_e=\bar{n}_i=1$), now dominating emitter photons.  Here, 
direct measurement fails. 
After 50 measurements, the confidence $c_e=0.58$, barely better than a random guess. 
This sensitivity to background noise is what motivates 
 the use of quantum interference.

Brute force simulation of measurement trajectories is computationally intensive.
To allow more extensive analysis and obtain analytic understanding we employ an alternative, approximate approach
 valid for large $N$.
 We first observe that  $\ln \Lambda =  \sum_{m=1}^N \ln \lambda_{jk}^{(m)}$
 is a sum of independent random variables $ \ln \lambda_{jk}^{(m)}$,
with mean $\mu_{\ln \! \lambda}(\xi) \! = \! \sum_{j,k=0}^\infty {p}_{jk} \, \ln \lambda_{jk}$ 
and variance $\sigma_{\ln \! \lambda}^2(\xi) \! = \! {  \sum_{j,k=0}^\infty {p}_{jk} \, (\ln \lambda_{jk} )^2 - \mu_{\ln \! \lambda}^2}$.
 By the Central Limit Theorem, for sufficiently large $N$ the distribution of $\ln \Lambda$ can therefore be approximated as normal with mean $\mu_{\ln \! \Lambda} = N \mu_{\ln \! \lambda}$ and standard deviation $\sigma_{\ln \! \Lambda} = \sqrt{N} \sigma_{\ln \! \lambda}$. 
$\Lambda$ then approaches the log-normal distribution~\cite{Wilks} 
\begin{equation}
p(\Lambda, \xi) = \frac{1}{  \sqrt{2 \pi} \, \Lambda \, \sigma_{\ln \! \Lambda}}  \exp \left ( - \frac{(\ln \Lambda - \mu_{\ln \! \Lambda})^2}{2 \sigma^2_{\ln \! \Lambda}} \right ). \label{log_norm_dist}
\end{equation}

To validate this approximation, we simulate 
 probability distributions of $\Lambda$ for the case of direct measurement with moderate noise ($\bar{n}^{\text{direct}}_n=1$). The
 simulated distributions approach log-normal as $N$ increases, with very good agreement once $N\geq 50$. The higher noise typical when
 emitter detection is challenging will randomise the distribution faster, leading to convergence at even lower measurement numbers.
  As a further test,  we use $p(\Lambda,\xi)$ together with the photon statistics of direct measurement (see Supplementary Material~\cite{SI}) to calculate the mean of $P_e$ for the same parameters as our earlier measurement trajectory simulations (Fig.~\ref{fig:diagram}{\bf c}, dashed lines).
  In the low noise case (blue) good agreement is observed with the mean from the simulations (solid lines) at all measurement numbers. In the higher noise case (red) the agreement is so strong that the curves cannot be distinguished.  

The condition $P_e>0.5$ for deciding that an emitter is present corresponds to the likelihood ratio condition $\Lambda <1$ (see  Eq.~(\ref{eq_Pe})). Thus, the chance that we correctly decide that an emitter is present in a given experiment is just
 $c_e  =  \int_{0}^1 p(\Lambda, \xi) \, d \Lambda$. Similarly, if the emitter is not present, $c_{\bar{e}} = \int_1^{\infty} p(\Lambda, 0) \, d \Lambda$. Since either possibility is equally likely, our overall confidence in a correct decision is $c=(c_e + c_{\bar{e}})/2$.
The properties of the log-normal distribution permit the  analytical solution:
\begin{equation}
c =  \frac{1}{2} + \frac{1}{4} \left [ \text{erf} \left (\! \! \sqrt{\frac{N}{2}} \frac{\mu_{\ln \! \lambda}(0)}{\sigma_{\ln \! \lambda}(0)} \! \right ) \! - \text{erf} \left (  \! \!  \sqrt{\frac{N}{2}} \frac{\mu_{\ln \! \lambda}(\xi)}{\sigma_{\ln \! \lambda}(\xi)} \! \right )  \! \right ] \!.\label{total_confidence}
\end{equation}
We see that to maximize confidence,
the magnitude of the error function arguments should be large and of opposite sign.
 For a fixed number of measurements, this can only be achieved by maximising the ratio of the mean and standard deviation of the log-likelihood, each of which are determined by the relevant photon count probability distribution (Eq.~(\ref{P_coh}) for extended HOM, Eq.~(S13) in Ref.~\cite{SI} for direct measurement).
 
{\bf Speed-up from HOM:} We are ultimately interested in the number of measurements required to confidently decide whether the emitter is present, and therefore how much faster -- or slower -- HOM interference is. 
This can be found by numerically inverting Eq.~(\ref{total_confidence}) to find $N$ at a given confidence level.
We choose two-sigma confidence, $c=0.954$, so that the chance of a correct answer is 95.4\%. 

\begin{figure}
	\includegraphics[width=\columnwidth]{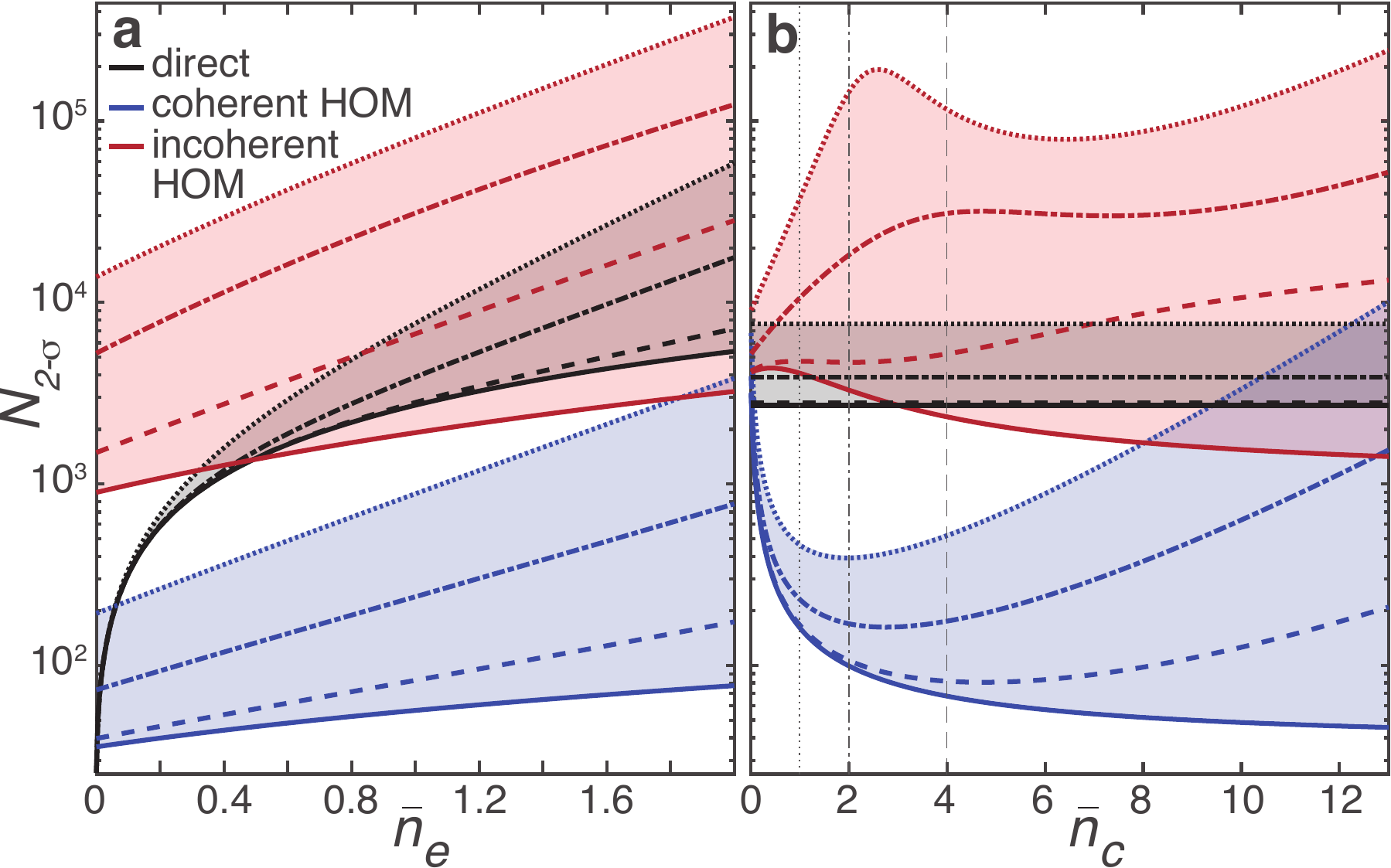}
	\caption{{\bf a}\&{\bf b} Number of measurements required for two-sigma confidence as a function of $\bar{n}_e$ and $\bar{n}_c$, respectively. Black: direct measurement; blue: coherent HOM; red: incoherent HOM. Solid curves: saturation-free detectors.  Dashed, dot-dashed, and dotted curves: saturation at $t=(4,2,1)$, respectively. $\eta=0.9$. 
	 {\bf a},  $\bar{n}_c = 6$; {\bf b}, $\bar{n}_e = \bar{n}_i=1$.
	\label{fig:N}  }
\end{figure}

Figure~\ref{fig:N} explores the influence of background noise and coherent photon number on the number of measurements needed to reach two-sigma confidence for direct measurement (black traces), coherent HOM (blue traces) and incoherent HOM (red traces). 
To ensure fair comparison, identical $\bar{n}_e$, $\bar{n}_i$ and $\eta$ are used for all protocols. 
We model both saturation-free photon number resolving detectors (solid lines), that can count an arbitrary number of photons, and detectors that saturate when reaching a threshold $t$ of four, two or one photons  (dashed lines), with $t=1$ corresponding to simple photon counting.
For saturating detectors, we truncate the photon count probability distributions at the threshold $t$ and cumulate the probabilities of higher photon counts into those at that threshold.

From Fig.~\ref{fig:N}{\bf a} we see that in all cases
the number of  measurements required to reach two-sigma confidence increases monotonically with background noise. 
Fig.~\ref{fig:N}{\bf b} shows that for both coherent and incoherent HOM the number of required measurements decreases with coherent state brightness, $\bar{n}_c$~\cite{exception}.
Naturally, detector saturation decreases the information content of each measurement, increasing the number required to reach two-sigma confidence in all cases.  For both forms of HOM it also degrades the performance at high $\bar{n}_c$. With saturation, the optimum coherent photon number is finite and non-zero for coherent HOM. For incoherent HOM it is zero, meaning that a vacuum local oscillator is more effective than a coherent one for the saturation thresholds investigated.

Even with simple photon counters ($t=1$),
coherent HOM outperforms direct measurement by more than an order of magnitude for appropriately chosen coherent state brightness, except where background noise is very low (Fig.~\ref{fig:N}{\bf a}). By contrast, incoherent HOM is significantly degraded,  only outperforming direct measurement for saturation-free detectors and when background noise or coherent photon number is sufficiently high.

\begin{figure}
	\includegraphics[width=\columnwidth]{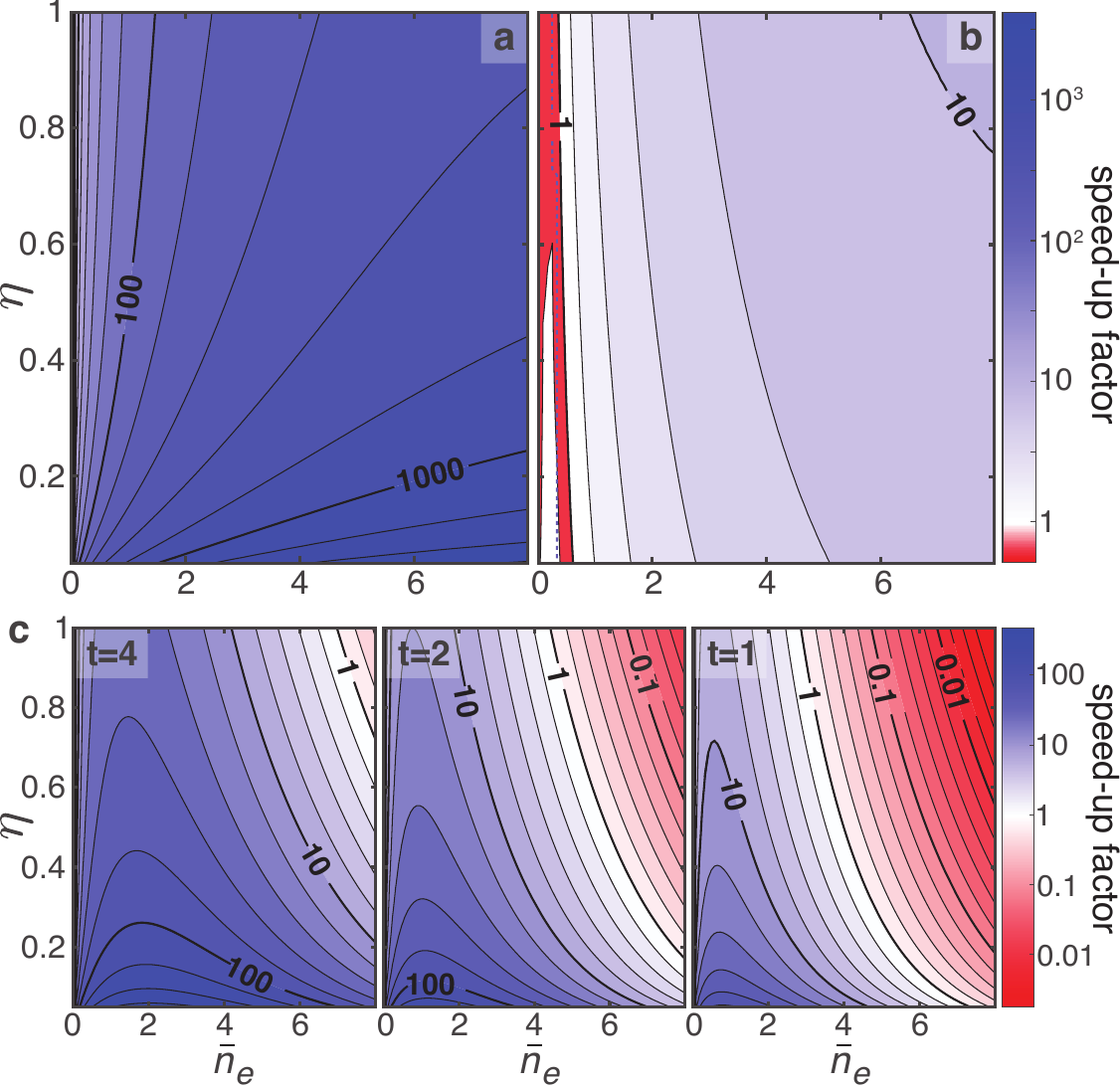}
	\caption{Speed-up factor as a function of detection efficiency and background noise, at optimum $\bar{n}_c$ for ({\bf a}) coherent, ({\bf b}) incoherent, and ({\bf c}) saturating-coherent HOM. Blue dot-dashed line in {\bf b}: transition between zero and bright optimal coherent state inputs. 
	\label{fig:speed}  } 
\end{figure}

We define the {\it speed-up factor} as the ratio of the number of measurements required for two-sigma confidence with direct measurement and HOM interference. Fig.~\ref{fig:speed}{\bf a}\&{\bf b} plot this for coherent and incoherent HOM as a function of detection efficiency and background photon counts for saturation-free detectors, optimised over coherent state brightness. For coherent HOM, the speed-up asymptotes to a maximum at high $\bar{n}_c$. For incoherent HOM, the same behaviour occurs at high background noise levels, but at low levels the optimum switches to a vacuum input (blue dot-dashed line, Fig.~\ref{fig:speed}{\bf b}). 

Fig.~\ref{fig:speed}{\bf a} shows that coherent HOM enables several orders-of-magnitude speed-ups over a wide parameter range, allowing emitter detection even with noise levels far higher than the emitter signal. Strikingly, the speed-up improves as both noise and detection losses increase; i.e., direct measurement is more fragile than coherent HOM. Incoherent HOM also speeds up emitter detection over a wide parameter range, though by a smaller margin -- at most just over an order of magnitude across the parameter range surveyed (Fig.~\ref{fig:speed}{\bf b}). 
 Like coherent HOM, its speed-up improves as noise increases but, conversely, degrades
(at moderate-and-above noise levels) with increasing detection loss.
 It is inferior to direct measurement at low noise levels and high efficiencies (Fig.~\ref{fig:speed}{\bf b}, red region). 

Figure~\ref{fig:speed}{\bf c} shows the speed-up from coherent HOM as a function of detection efficiency and background noise for saturating detectors.
In this case, the optimum $\bar{n}_c$ varies strongly depending on the level of noise, saturation threshold, and detection efficiency (Fig. 4, Supplementary Material~\cite{SI}). The speed-up improves as background noise increases, but degrades at higher noise levels due to saturation.
When operating near the optimum noise levels, order-of-magnitude speed-ups remain possible, even for photon counters ($t=1$).  The improvement in speed-up with decreasing detection efficiency remains, illustrating the robustness of HOM-based emitter detection. Indeed, speed-ups of several orders of magnitude are possible even with $\eta < 0.1$. By contrast, even with a four-photon saturation threshold,
 incoherent HOM is inferior to direct measurement.

{\bf Conclusion and outlook:} We have shown that extended HOM interference combined with Bayesian inference can detect quantum emitters orders-of-magnitude faster than direct detection, robust to imperfections such as background noise, detector loss and detector saturation. The method works for both coherent superpositions and mixed states.

Many practical quantum information protocols are expected to require large arrays of quantum emitters~(e.g., Refs.~\cite{psiquantum2025manufacturable,kielpinski2002architecture,azuma2023quantum}), with increasing interest in 
employing coherent superposition states of zero and one photon (e.g., Refs.~\cite{loredo2019generation,polacchi2023quantum,bozzio2022enhancing,renema2020simulability,drahi2021entangled}).
Our protocol could prove useful in rapidly assessing the quality of these emitters. It may also be adaptable to enable robust characterisation of photonic quantum states, thereby improving the performance of measurement-based quantum computation~\cite{bartolucci2023fusion,nielsen2006cluster}.

Quantum emitters are central to a wide range of microscopy techniques, such as fluorescence, second-harmonic generation, and Raman microscopy. Often, these microscopes are plagued by background noise~\cite{mandracchia2020fast,kubow2011reducing}. Resolving the locations of emitters against this noise, at high speed, and with the minimal optical intrusion onto fragile biological specimens is a central challenge~\cite{deschout2014precisely,rust2006sub,vicidomini2018sted,li2020adaptive,casacio2021quantum}.
Our protocol has the potential to do this. The combination of robustness to loss and background noise, along with prior success in applying HOM interference to spectroscopy~\cite{eshun2021investigations}, microscopy~\cite{ndagano2022quantum,li1997femtosecond,lyons2023fluorescence}, and camera-based imaging~\cite{jachura2015shot,ndagano2022quantum}, evidence its experimental feasibility. 

\begin{acknowledgments}
W.P.B. acknowledges valuable discussions with Tim Ralph and Kyle Clunies-Ross. This work was supported by the Australian Research Council Centre of Excellence in Quantum Biotechnology (QUBIC, CE230100021), the Air Force Office of Scientific Research (AFOSR, grants FA9550-22-1-0047 \& FA9550-20-1-0391), and a Chan-Zuckerberg Initiative Deep Tissue Imaging grant (DTI2-0000000182).
\end{acknowledgments}

\bibliography{Ref}

\begin{thebibliography}{42}
\bibitem [{Wil()}]{Wilks}%
  \BibitemOpen
  \href@noop {} {}\bibinfo {note} {Note: by Wilks' Theorem, a $\chi^2$
  distribution is generally more accurate at low measurement numbers, but
  agrees with Eq.~(\ref{log_norm_dist}) in our high-measurement-number
  limit.}\BibitemShut {Stop}%
\end{thebibliography}

\clearpage
\onecolumngrid  

\begin{center}
\textbf{\large Supplementary Material for: ``Large speed-up of quantum emitter detection via quantum interference''}

\vspace{1.5em}

\textbf{Warwick P. Bowen}\\[0.5em]

{\it Australian Research Council Centre of Excellence in Quantum Biotechnology, School of Mathematics and Physics, University of Queensland, St Lucia, Queensland 4072, Australia}\\[1.5em]

\parbox{0.85\linewidth}{
\textbf{Abstract:} 
These supplementary notes go through detailed derivations of the photon number statistics of direct measurement, coherent Hong-Ou-Mandel interferometry and incoherent Hong-Ou-Mandel interferometry. They introduce the approach used to Bayesian inference and to determine the minimum number of measurements required to achieve a desired level of confidence, including derivation of the log-normal distribution of the likelihood ratio. Also included is a plot of the optimal coherent photon number for coherent Hong-Ou-Mandel interferometry with saturating photon resolving detectors, complimenting Fig.~3{\bf c} of the main text.
}
\end{center}

\vspace{1.5em}

\setcounter{equation}{0}
\setcounter{figure}{0}
\setcounter{table}{0}
\setcounter{section}{0}
\renewcommand{\theequation}{S\arabic{equation}}
\renewcommand{\thefigure}{S\arabic{figure}}
\renewcommand{\thesection}{S\arabic{section}}
\renewcommand{\thetable}{S\arabic{table}}

\pagenumbering{arabic}          
\setcounter{page}{1}            
\renewcommand{\thepage}{S\arabic{page}}  

\section*{S1. Quantum emitter detection with direct measurement}

\begin{figure}[h!]
	\begin{center}
	\includegraphics[scale=0.4]{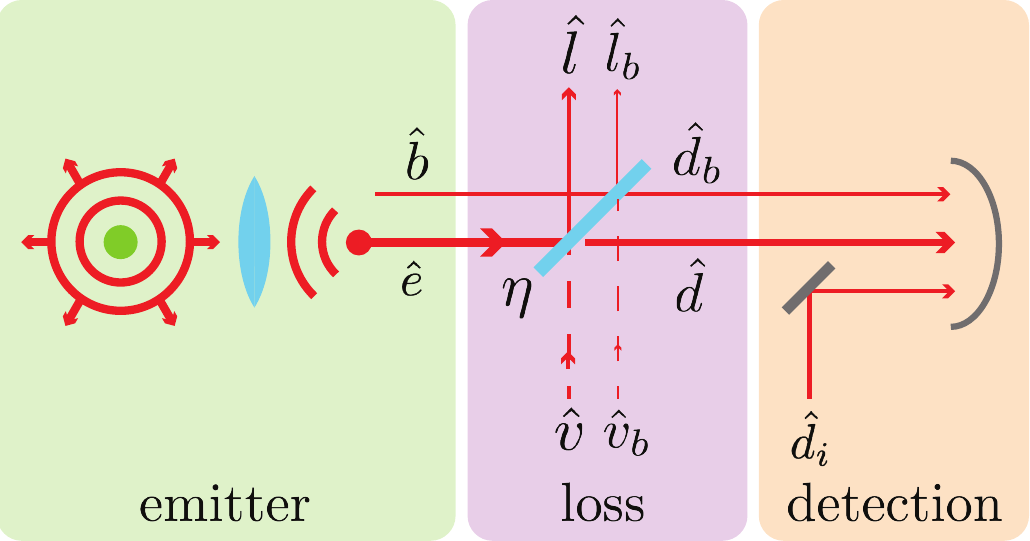}
	\caption{Direct measurement of emitter emission, including loss and background counts both generated in the sample and in the detector. $\eta$: detection efficiency, $\hat e$: annihilation operator for emitter output mode, $\hat b$: annihilation operator for background emission mode at emitter, $\hat v$ annihilation operator for vacuum input into emitter mode due to inefficiency, $\hat v_b$: annihilation operator for vacuum input to background emission mode due to inefficiency, $\hat d_i$: annihilation operator to model background photocurrent noise at the detector, $\hat d$: annihilation operator for detected field in emission mode, $\hat d_b$: annihilation operator for detected field in background emission mode, $\hat l$: annihilation operator for output loss mode, $\hat l_b$ annihilation operator for output background emission mode.	
	 }
	\label{fig:direct}%
	\end{center}
\end{figure}

Figure~\ref{fig:direct} illustrates the process of direct measurement of the emitted field from a quantum emitter, including various imperfections. The emitter is shown in the green box, with emitted light collected and focussed onto a photon number resolving detector. The detection process involves inefficiency, which is modelled by beam-splitter loss $\eta$, and also introduces electronic noise. Background noise generated near the emitter, for instance by autofluorescence~\cite{waters2009accuracy}, is included and assumed to be in an orthogonal spatiotemporal mode to the emitted emission.

\subsection{a. Photon statistics}
\label{direct_phot_stat}

We model the emitter as weakly pumped so that it emits a superposition state of zero and one photon
\begin{equation}
| e \rangle  =  \sqrt{1-\xi} |0 \rangle + \sqrt{\xi} |1 \rangle =  \left ( \sqrt{1-\xi} + \sqrt{\xi} \hat e^\dagger \right ) | 0 \rangle 
\end{equation}
where $\xi$ is the probability of photon emission and $\hat e^\dagger$ is a creation operator adding a photon from the spatiotemporal mode of the emitter. The background emission and detection noise are modelled as random noise processes in independent detection modes to the emitter emission. On combined detection, these noise sources introduce a Poissionian background of photon clicks which we include later. All other modes are initially in vacuum states.

We represent the initial state of the  emission mode, the input vacuum noise it interferes with, the output detection mode, and the output loss mode as 
\begin{equation}
|\psi_{\rm in} \rangle = |e \rangle \otimes |v \rangle \otimes |d \rangle \otimes | l \rangle = | e, v, d, l \rangle = \left ( \sqrt{1-\xi} + \sqrt{\xi} \hat e^\dagger \right ) |0,0,0,0\rangle
\label{initial_state_direct}
\end{equation}

To determine the final state we propagate annihilation operators for each input mode through the system to obtain the output annihilation operators
\begin{eqnarray}
\hat d &=& \sqrt{\eta} \hat e + \sqrt{1-\eta} \hat v\\
\hat l &=& \sqrt{1-\eta} \hat e - \sqrt{\eta} \hat v.
\end{eqnarray}
Rearranging for the input operators we get
\begin{equation}
\hat e = \sqrt{\eta} \hat d + \sqrt{1-\eta} \hat v.\\
\end{equation}

Substitution into Eq.~(\ref{initial_state_direct}) yields the final state
\begin{equation}
|\psi_{\rm out} \rangle = \sqrt{1-\xi}  |0,0,0,0\rangle + \sqrt{\xi \eta} |0,0,1,0\rangle + \sqrt{\xi (1-\eta)} |0,0,0,1\rangle.
\label{final_state_direct}
\end{equation}
As expected, the probabilities of detecting an emitted photon in the detection mode $d$, generating one but losing it out the loss port, and never generating it at all are respectively,
\begin{eqnarray}
P_{1,0} &=& |\langle 0, 0, 1, 0 |\psi_{\rm out} \rangle|^2  = \xi \eta 
\label{p_d_direct} \\ 
P_{0,1} &=& |\langle 0, 0, 0, 1 |\psi_{\rm out} \rangle|^2  = \xi (1- \eta)\\
P_{0,0} &=& |\langle 0, 0, 0, 0 |\psi_{\rm out} \rangle|^2  = 1 - \xi,
\end{eqnarray}
where the first subscript labels the number of photons in the detection mode, and the second labels the number of photons in the loss mode. Tracing over the unmeasured output loss mode, we find the probabilities of detecting no or one emitter photon in the detection mode
\begin{eqnarray}
\bar P_{{\rm emitter}, 0} &=& 1- \xi \eta 
\label{p_d_direct_0} \\ 
\bar P_{{\rm emitter}, 1} &=&  \xi \eta,
\label{p_d_direct_1} 
\end{eqnarray}
where the subscript represents the number of detected emitter photons and, throughout this supplement, we use the accent `bar' on $\bar P$ to indicate that we have traced over undetected modes.

As described above, we treat stray photon counts that originate from the vicinity of the emitter and within the detectors as uncorrelated processes in orthogonal spatiotemporal modes to the emitter mode. Each process is described by a Poissonian photon number distribution with mean number of detected photons $\bar n_j$, where $j \in \{ e, i \}$. $\bar{n}_e$ is the background noise introduced in the emitter location, and $\bar{n}_i$ is the detector electronic noise. Propagating the noise photons produced near the emitter to the detector maintains the Poissonian distribution, but with reduced mean of $\eta \bar n_e$. The total noise photon distribution is therefore described by a single Poissonian distribution with overall mean $\bar n_{n} = \eta \bar n_e + \bar n_i$, and probability distribution
\begin{equation}
\bar P_{{\rm noise},k} = e^{-\bar n_n} \frac{\bar n_n^k}{k!}, \label{poisson_direct}
\end{equation}
where $k$ is the total number of detected stray photons.

Since the stray photons are treated as being emitted into orthogonal spatiotemporal modes to the emitter mode, the combined probability distribution of detecting a total of $k$ photons is given by the convolution of Eq.~(\ref{poisson_direct}) with the probabilities of detecting or not detecting an emitted photon given in Eqs.~(\ref{p_d_direct_0})~and~(\ref{p_d_direct_1}). Performing this convolution, we find
\begin{equation}
\bar P_k = \sum_{i+j=k}  \bar P_{{\rm emitter}, i} \, \bar P_{{\rm stray},j} =
 e^{-\bar n_n} \frac{\bar n_n^k}{k!} \left (1 - \eta \xi + \eta \xi \frac{k}{\bar n_n} \right ). \label{direct_det_dist}
\end{equation}
This final detected photon number distribution has a Poissonian pre-factor, with the term in parenthesis modifying the statistics due to the non-Poissonian character of the quantum emission. This is characterised by two parameters, the mean intensity (in photon number units) of the detected quantum emission $\xi \eta$ and the mean detected noise photon number $\bar{n}_n$.

\section*{S2. Bayesian estimation}
\label{bayes_direct}

The photon number distribution in Eq.~(\ref{direct_det_dist}) provides all of the statistical information available about the detected photon counts, given the strength of emission, detection efficiency, and background counts. Using it, together with the distribution in the case that no emitter exists (setting $\xi=0$), we are able to apply Bayesian inference to determine the likelihood that an emitter existed given a set of outcomes from a series of $N$ measurements.

In Bayesian inference, the probability of some event $A$ after a measurement $B$ is given by~\cite{good1989weight}
\begin{equation}
P(A|B) = \frac{P(B|A) P_{\rm prior} (A) }{P(B)},
\end{equation}
where $P_{\rm prior}(A)$ is the probability of the event prior to the measurement, $P(B)$ is the probability of the measurement outcome, and $P(B|A)$ is the conditional probability of the measurement outcome were the event $A$ to be true. In Bayesian inference from a sequence of $N$ measurements, after each measurement the prior for the next measurement is updated as  $P_{\rm prior} (A) \rightarrow P(A|B)$, or, including the sequence of measurements explicitly as
\begin{equation}
P(A|B_1,B_2,B_3...B_N) = P_{\rm prior} (A)  \prod_{j=1}^N \frac{P(B_j|A) }{P(B_j)}.
\end{equation}

In our case, the events that we wish to distinguish are the presence or absence of an emitter, which we denote $A \in \{ e, \bar e\}$, with $\bar e$ being the no-emitter case. We then have
\begin{eqnarray}
P(e) = P_{\rm prior} (e)  \prod_{j=1}^N \frac{P(k_j| e) }{P(k_j)}\\
P(\bar e) = P_{\rm prior} (\bar e)  \prod_{j=1}^N \frac{P(k_j| \bar e) }{P(k_j)},
\end{eqnarray}
where for succinctness we have replaced the label $A|B_1,B_2,B_3...B_N$ on the left-hand-side with $e$ and $\bar e$ to denote the overall Bayesian probabilities after all measurements, and the outcomes $B_j$ are given by the detected photon number $k_j$ from each run of the measurement. [Note: in the main text we use the alternative notation $P_e(\xi)$ and $P_e(0)$ to denote the Bayesian probability distribution in the presence and absence of an emitter.]
Since there are only two possible scenarios $P(e)  + P(\bar e) =1$. Using this, the probability of the measurement outcomes $P(k_j)$ can be cancelled from the expressions for the probabilities. E.g., the  probability that the emitter is present can be re-expressed as
\begin{eqnarray}
P(e)  &=& \frac{P(e) }{P(e) +P(\bar e) } \nonumber \\
&=& \frac{P_{\rm prior} (e) }{P_{\rm prior} (\bar e) } \times \left ( \frac{\prod_{j=1}^N P(k_j| e)}{\prod_{j=1}^N P(k_j| e)+\prod_{j=1}^N P(k_j| \bar e) } \right )\nonumber \\
&=&  \frac{P_{\rm prior} (e) }{P_{\rm prior} (\bar e) } \times \left (  \frac{1}{1+  \prod_{j=1}^N \lambda_j } \right) \nonumber \\
&=&  \frac{P_{\rm prior} (e) }{P_{\rm prior} (\bar e) } \times \left (  \frac{1}{1+  \Lambda} \right),
\end{eqnarray}
where 
\begin{equation}
\lambda_j  = \frac{P(k_j| \bar e) }{P(k_j| e)}, \label{lambda_j_eqn}
\end{equation}
is referred to as the {\it likelihood ratio}~\cite{good1989weight} on an individual measurement and $\Lambda$ is the overall likelihood ratio after the full sequence of measurements. A likelihood ratio above (below) 1 indicates that the emitter is more likely to be present (absent), with a larger (smaller) value indicating a higher level of confidence.

In the paper we set a prior probability $P_{\rm prior} (e) = 0.5$, i.e., we assume that we have no prior information about the presence of the emitter. Our task in determining the Bayesian probability of the presence of the emitter then reduces to determining $\Lambda$:
\begin{eqnarray}
P(e) &=&  \frac{1}{1+\Lambda} \label{P_eqnB} \\
P(\bar{e})  &=& 1 - P(e) =  \frac{1}{1+\Lambda^{-1}}. \label{P_bar_eqnB}
\end{eqnarray}

We wish to determine the distribution of $P(e)$ values that could result from a series of $N$ measurements, or {\it measurement trajectories},  both in the case where the emitter is present and when it is absent.
The most direct way to do this is to perform a series of simulations of photon count outcomes, calculate trajectories of $P(e)$ for each, and thus determine the distribution. However, this is computationally intensive, requiring many simulations to achieve high accuracy. For this reason we use an alternative approach that provides approximate probability distributions 
 in the limit of large $N$ without requiring simulations of measurement trajectories. The key is to determine an approximate expression for  the probability distribution of the overall likelihood ratio $\Lambda$. 

We take the natural logarithm of $\Lambda$ to define the new variable
\begin{equation}
y = \ln  \Lambda = \sum_{j=1}^N \ln \lambda_j= \sum_{j=1}^N x_j,
\end{equation}
which we see can be expressed as a sum of the random variable $x_j = \ln \lambda_j$ with mean $\mu_x = \langle  \ln \lambda_j \rangle$ and standard deviation $\sigma_x = [ \langle  (\ln \lambda_j)^2 \rangle  - \mu_x^2]^{1/2}$. By the Central Limit Theorem, for sufficiently large $N$, the distribution of $y$ can be approximated as a normal distribution with mean $\mu_y$ and standard deviation $\sigma_y$
\begin{eqnarray}
\mu_y &=& N \mu_x \label{mu_y} \\
\sigma_y &=& \sqrt{N} \sigma_x \label{sigma_y}.
\end{eqnarray}
Therefore, in this limit $\Lambda = e^y$ has the log-normal distribution parametrised by the same mean and variance~\footnote{Note: while $\mu_y$ and $\sigma_y$ parametrise the log-normal distribution, they should not be interpreted as its mean and standard deviation.}
\begin{equation}
p(\Lambda) = \frac{1}{\Lambda \sigma_y \sqrt{2 \pi}} \exp \left ( - \frac{(\ln \Lambda - \mu_y)^2}{2 \sigma_y^2} \right ). \label{log_norm_dist_supp}
\end{equation}
Note that Wilks' Theorem generally provides a more accurate distribution at low measurement numbers -- the $\chi^2$ distribution~\cite{algeri2020searching}. However, this agrees with Eq.~(\ref{log_norm_dist_supp}) in the large measurement number limit considered in this paper.

$\mu_y$ and $\sigma_y$ can be determined from $\mu_x$ and $\sigma_x$ using Eqs.~(\ref{mu_y})~and~(\ref{sigma_y}), while $\mu_x$ and $\sigma_x$ can be calculated from the photon counting probability distribution of a given protocol $\bar P_k(e)$ as
\begin{eqnarray}
\mu_x(e) &=& \sum_{k=0}^\infty P_k(e) \, x_k \label{mu_x_e}\\
\sigma_x^2(e) &=& { \sum_{k=0}^\infty P_k(e) \, x_k^2 - \mu_x^2(e)}, \label{sigma_x_e}
\end{eqnarray}
where $x_k = \text{ ln} \, \lambda_k$ and the sum is over all possible measurement outcomes.  
For the direct measurement case considered in the previous Section,  $\bar P_k(e)$ is given in Eq.~ (\ref{direct_det_dist}). The argument $e$ in $\bar P_k(e)$ is introduced to label that this probability distribution applies when the emitter is present. The probability distribution without emitter present $\bar P_k(\bar{e})$ is obtained by  setting $\xi=0$ in  Eq.~ (\ref{direct_det_dist}).
 Using both probability distributions in Eq.~(\ref{lambda_j_eqn}), we find the likelihood ratio of each possible measurement outcome $k$  for direct measurement
\begin{equation}
\lambda_k = \frac{\bar{P}_k(\bar e) }{\bar{P}_k(e)} = \left (1 - \eta \xi + \eta \xi \frac{k}{\bar n_n} \right )^{-1}.
\end{equation}

Using this result we can evaluate each of Eqs.~(\ref{mu_x_e})--(\ref{sigma_x_e}) for the mean and standard deviation of $x$. We find that they are given by
\begin{eqnarray}
\mu_x(e) &=& - e^{-\bar n_n}  \sum_{k=0}^\infty \frac{\bar n_n^k}{k!} \left (1 - \eta \xi + \eta \xi \frac{k}{\bar n_n} \right )  \, \text{ln} \, \left (1 - \eta \xi + \eta \xi \frac{k}{\bar n_n} \right ) \label{mean1}\\
\sigma_x^2(e) &=& e^{-\bar n_n}  \sum_{k=0}^\infty \frac{\bar n_n^k}{k!} \left (1 - \eta \xi + \eta \xi \frac{k}{\bar n_n} \right ) \left [ \, \text{ln} \, \left (1 - \eta \xi + \eta \xi \frac{k}{\bar n_n} \right ) \right ]^2  - \mu_x(e)^2,
\end{eqnarray}
in the presence of an emitter and 
\begin{eqnarray}
\mu_x(\bar{e}) &=& - e^{-\bar n_n}  \sum_{k=0}^\infty \frac{\bar n_n^k}{k!}  \, \text{ln} \, \left (1 - \eta \xi + \eta \xi \frac{k}{\bar n_n} \right )\\
\sigma_x^2(\bar{e}) &=& e^{-\bar n_n}  \sum_{k=0}^\infty \frac{\bar n_n^k}{k!}  \left [ \, \text{ln} \, \left (1 - \eta \xi + \eta \xi \frac{k}{\bar n_n} \right ) \right ]^2  - \mu_x(\bar{e})^2, \label{sigma2}\
\end{eqnarray}
 in its absence.

\subsection{a. Validation of the log-normal distribution}

We validate the use of a log-normal probability distribution for $\Lambda$  through comparison with simulations. First, we simulate the probability distribution itself after sequences of measurements of different length, with results shown in Fig.~\ref{validate_CLT}. We simulate trajectories for three different numbers of measurements, $N=\{10,30,50\}$ and for the parameters given in the Figure caption. In each case, 50,000 trajectories are simulated and the final distributions of $y= \ln  \Lambda$ are calculated both when an emitter is present (blue histograms) and absent (red histograms). The simulations are compared to analytic curves based on the log-normal distribution, with no fitting parameters.  When an emitter is present (absent), the histograms shift negative (positive), as expected ($y < 0$ when the Bayesian likelihood is that the emitter is present). The shorter simulations show discrete probability spikes (Fig.~\ref{validate_CLT}{\bf a}\&{\bf b}). This is due to the discreteness of  the measurement outcomes -- when only a small number of measurements are taken there are only a few likely combinations of measurement outcomes. As the measurement number increases (Fig.~\ref{validate_CLT}{{\bf c}), the histograms of $y$ appear more continuous and approach a normal distribution in good agreement with the analytic curves. This indicates that, even for relatively small series of measurements ($N=50$), the log-normal distribution is a suitable approximation for  $\Lambda$ (since $y=\ln \Lambda$). Since we are expressly interested in scenarios where it is {\it hard} to detect the emitter, we expect many more measurements to be required and therefore the log-normal distribution to be highly accurate.
\begin{figure}[h!]
	\begin{center}
	\includegraphics[scale=0.6]{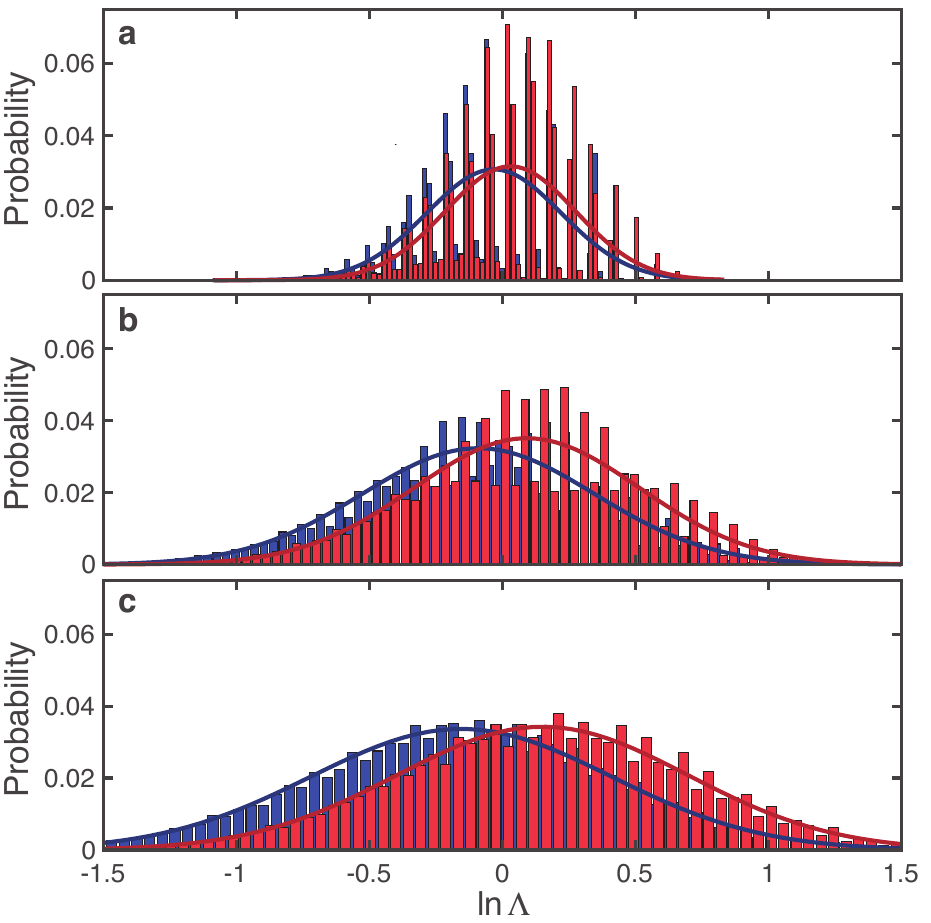}
	\caption{Probability distribution of the log of the likelihood ratio ($y=\ln \Lambda$) when the emitter is present (blue) and absent (red), as a test of the central limit theorem approximation. The offset between distributions with and without an emitter is indicative of the resolvability of the presence of the emitter. {\bf a.} Probability distributions after $N=10$ measurements.  {\bf b.} Probability distributions after $N=30$ measurements.  {\bf c.} Probability distributions after $N=50$ measurements. For all simulations $\bar{n}_n=1$, $\xi=0.1$, and $\eta=0.8$, and 50,000 trajectories were used.
	 As $N$ increases the distributions approach Gaussians with mean and standard deviation given by $\mu_y$ and $\sigma_y$. The simulations agree well with theory for higher $N$ values (curves), with $\mu_y$ and $\sigma_y$ calculated from Eqs.~(\ref{mu_y}),~(\ref{sigma_y})~and~(\ref{mean1}-\ref{sigma2}) with no fitting parameters. The spikes in the probability distributions, especially noticeable at low $N$ are because there are a discrete set of possible outcomes for $\Gamma$. The discreteness gets washed out as $N$ increases. }
	 \label{validate_CLT}
	\end{center}
\end{figure}

Second, we use the  log-normal probability distribution to calculate the mean of the Bayesian probability $\langle P(e) \rangle$, and compare this to the mean of the simulations given in Fig.~1{\bf c} of the main paper.
$\langle P(e) \rangle$ is calculated as
\begin{equation}
\langle P(e) \rangle = \int_0^\infty  \frac{p(\Lambda)}{1+\Lambda} \, d \Lambda. \label{mean_P}
\end{equation}
Using Eq.~(\ref{log_norm_dist_supp}) and making the substitution $y=\text{ln} \, \Lambda$ we obtain the convenient to evaluate expression
\begin{equation}
\langle P(e) \rangle =   \frac{1}{\sigma_y \sqrt{2 \pi}}  \int_{-\infty}^\infty  \frac{1}{1+e^y} \exp \left ( - \frac{(y-\mu_y)^2}{2 \sigma_y^2} \right ) dy. \label{mean_P_in_terms_of_y}
\end{equation}
Using this expression, we determine $\langle P(e) \rangle$ for direct measurement, as a function of the number of measurements and with the same parameters as used for the simulations in Fig.~1{\bf c} in the main text. The results are plotted on the figure, showing good agreement with the mean of  $P(e)$ from the set of simulated trajectories even for very low numbers of measurements. The robustness of the approximation, even when the probability distribution of $\Gamma$ is decidedly not log-normal (such as in Fig.~\ref{validate_CLT}{\bf a}), likely arises because the deviations in the distribution average out to some extent when calculating the mean of $P(e)$.

\subsection{b. Number of measurements required to reach desired confidence}
\label{num_meas_sec}

Let us say that we wish to be able to register the presence or absence of the emitter with some level of confidence $c(e)$ (or $c(\bar{e})$), where $c$ tells you the probability that your decision is correct given all possible measurement trajectories. We need to decide how to choose a decision (whether the emitter is present or not), given the Bayesian probability $P(e)$ that results from the set of measurements. The natural choice is to decide that the emitter exists if $P(e)>0.5$ and doesn't exist else-wise. From Eq.~(\ref{P_eqnB}) we see that the transition occurs at $\Lambda=1$, with $P(e)>0.5$ if $0<\Lambda <1$ and $P(e)<0.5$ is $1<\Lambda<\infty$. Thus, we can determine our level of confidence in the presence and absence of the emitter from the probability distribution of $\Lambda$ as
\begin{eqnarray}
c(e) & = & \int_{0}^1 p(\Lambda | e) \, d \Lambda 
 \label{ce}\\
c(\bar{e}) &=& \int_1^{\infty} p(\Lambda |\bar{e}) \, d \Lambda. \label{cbare}
\end{eqnarray}
Here, we have made the replacement $p(\Lambda) \rightarrow \{ p(\Lambda | e),p(\Lambda | \bar{e}) \}$ to explicate that $p(\Lambda)$ depends on whether an emitter is present through its dependence on $\mu_y$ and $\sigma_y$ (see Eq.~(\ref{log_norm_dist_supp})). The properties of the log-normal distribution permit analytical solutions to these integrals in terms of the error function:
\begin{eqnarray}
c(e) & = & \frac{1}{2} \left [ 1 - \text{erf} \left ( \frac{\mu_y(e)}{\sqrt{2} \sigma_y(e)} \right ) \right ]
 \label{ceB}\\
c(\bar{e}) &=& \frac{1}{2} \left [ 1 + \text{erf} \left ( \frac{\mu_y(\bar{e})}{\sqrt{2} \sigma_y(\bar{e})} \right ) \right ]. \label{cbareB}
\end{eqnarray}

Given that we don't know if the emitter is present or not, we are ultimately interested in a combined confidence $c$ which takes into account both possibilities. Assuming either situation is equally likely,  consistent with our choice of Bayesian prior,
\begin{eqnarray}
c &=& \frac{c(e) +c(\bar{e})}{2}\\
&=& \frac{1}{2} + \frac{1}{4} \left [ \text{erf} \left (\sqrt{\frac{N}{2}} \frac{\mu_x(\bar{e})}{\sigma_x(\bar{e})} \right )  - \text{erf} \left (\sqrt{\frac{N}{2}} \frac{\mu_x(e)}{\sigma_x(e)} \right )   \right ], \label{total_confidence_supp}
\end{eqnarray}
where we have used Eqs.~(\ref{ceB})~and~(\ref{cbareB}), and substituted for $\mu_x$ and $\sigma_x$ using Eqs.~(\ref{mu_y})~and~(\ref{sigma_y}). This equation gives some intuition for what is required to confidently distinguish the presence or absence of an emitter: high confidence requires that the magnitude of the arguments of both error functions are large, but of opposite signs (positive for the case where the emitter is present and negative when it is not). This can be achieved both by having a log-likelihood distribution with large ratio of mean to standard deviation and by making a large number of measurements. We also note that if the ratios of mean to standard deviation with and without the emitter present are equal, the confidence reduces to one half -- the estimation reduces to a guess as expected since the distributions with and without the emitter are statistically identical.

Note, were, by Wilks theorem~\cite{algeri2020searching}, a $\chi^2$ log-likelihood distribution used instead of a log-normal distribution, the confidence could be similarly expressed in terms of the regularized lower incomplete gamma function. This, somewhat more complicated form is unnecessary here, since the log-normal distribution is accurate for all cases considered.

A particularly physically meaningful question is to ask is: given a set of parameters, how many measurements are required to achieve a certain level of overall confidence $c$. This can be determined numerically from Eq.~(\ref{total_confidence_supp}).
Throughout the paper, we set a two-sigma confidence level of $c=0.954$ and determine the number of measurements $N_{2\text{-}\sigma}$ required to reach this confidence.

\section*{S3. Quantum emitter detection with extended Hong-Ou-Mandel interference}

\begin{figure}[h!]
	\begin{center}
	\includegraphics[scale=0.5]{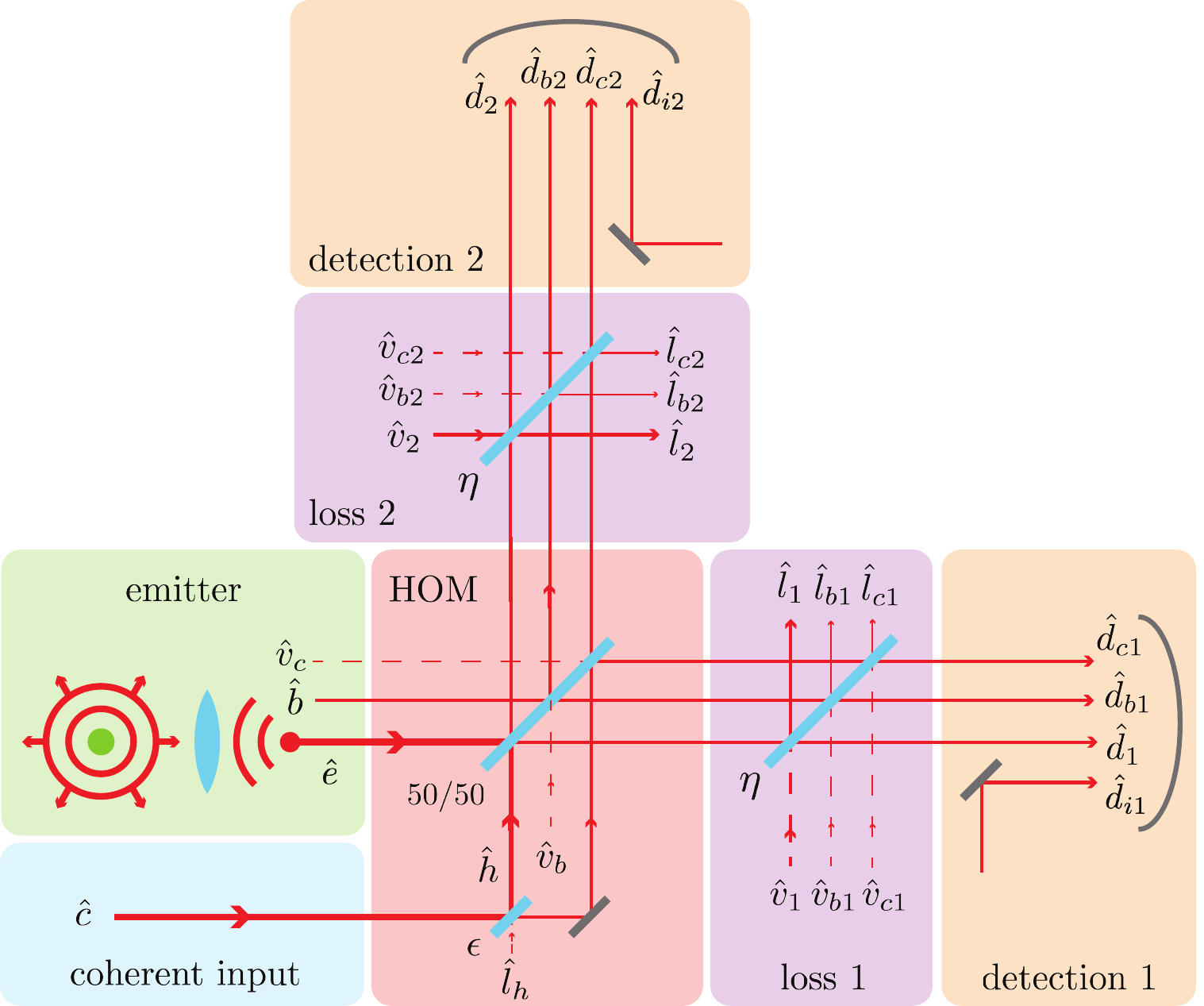}
	\caption{HOM detection of emitter emission, including loss and background counts both generated in the sample and in the detector, and including HOM mode matching inefficiency. $\eta$: detection efficiency, $\epsilon$: mode matching efficiency, $\hat e$: annihilation operator for emitter output mode, $\hat c$: annihilation operator for coherent input mode, $\hat b$: annihilation operator for background emission mode at emitter, $\hat v_j$ annihilation operators representing input vacuum modes, $d_j$ are annihilation operators representing detected output modes, $\hat l_j$ are operators representing undetected output loss modes, subscripts $\{1,2\}$ distinguish modes associated with each of the two HOM detectors. Subscripts $\{b,c,i \}$ denote background emission, coherent input, and detector photocurrent noise related modes.  }
	\label{fig:HOM}%
	\end{center}
\end{figure}

\subsection{a. Photon statistics}

We now extend the analysis to emitter detection via extended Hong-Ou-Mandel (HOM) interference. We consider the experimental configuration shown in Fig.~\ref{fig:HOM}. The input field from the emitter is described in the Heisenberg picture by the annihilation operators $\hat{e}$, as before, but now a second input field $\hat{c}$ is introduced. This interferes with the field from the emitter on a 50/50 beam splitter in a form of extended Hong-Ou-Mandel (HOM) interference~\cite{PhysRevA.86.063828, PhysRevA.105.013712, HOModd, rarity2005non,lyons2023fluorescence}. We include all of the imperfections already discussed for direct measurement, and also include imperfect overlap between the fields in the interferometer (modelled via the overlap factor $\epsilon$).  The two primary outputs of the interferometer, described by the annihilation operators $\hat{d}_1$ and $\hat{d}_2$, are detected on two separate photon number resolving detectors. Due to the imperfections, twelve other output fields must also be accounted for (see Fig.~\ref{fig:HOM}), six of which are detected and add noise to the measurements ($\hat{d}_{c1}$, $\hat{d}_{c2}$, $\hat{d}_{b1}$, $\hat{d}_{b2}$, $\hat{d}_{i1}$, $\hat{d}_{i2}$) and six are lost and must be traced over ($\hat{l}_1$, $\hat{l}_2$, $\hat{l}_{b1}$, $\hat{l}_{b2}$, $\hat{l}_{c1}$, $\hat{l}_{c2}$). Here, the subscripts $1$ and $2$ label the two arms of the interferometer; subscripts $i$, $b$ and $c$ respectively label terms associated with photocurrent noise, background emitter noise, and the coherent input; and $l$ and $d$ label fields that are lost and detected, respectively.

We first determine the photon statistics for only $\hat{d_1}$ and $\hat{d_2}$, introducing the effect of the six detected noise fields later.  As in the case of direct measurement (Section~S1),
we propagate the input annihilation operators through the apparatus to the detectors to obtain a set of linear equations for $\hat{d}_1$, $\hat{d}_2$, $\hat{l}_1$, and $\hat{l}_1$. Rearranging these equations in terms of the input annihilation operators we find
\begin{eqnarray}
\hat{e} &=& \sqrt{\frac{\eta}{2}} \left ( \hat{d}_1 + \hat{d}_2 \right ) + \sqrt{\frac{1-\eta}{2}} \left (\hat{l}_1 + \hat{l}_2 \right ) \label{HOM_e}\\
\hat{h} &=& -\sqrt{\frac{\eta}{2}} \left (\hat{d}_1 - \hat{d}_2 \right ) - \sqrt{\frac{1-\eta}{2}} \left (\hat{l}_1 - \hat{l}_2 \right )  \label{HOM_h}\\
\hat{v}_1 &=& \sqrt{1-\eta} \hat{d}_1 - \sqrt{\eta} \hat{l}_1\\
\hat{v}_2 &=& \sqrt{1-\eta} \hat{d}_2 - \sqrt{\eta} \hat{l}_2,
\end{eqnarray}
where, to simplify calculations, we leave the equation for the coherent input in terms of the component of the field that is perfectly overlapped with the emitter field $\hat{h} = \sqrt{\epsilon} \hat{c} + \sqrt{1-\epsilon} \hat{l}_h$, with $ \hat{l}_h$ being the annihilation operator associated with the vacuum input due to imperfect mode overlap. This works because the beam-splitter used to model mode-mismatch couples the input coherent state with a vacuum field, resulting in two uncorrelated coherent states at the beam-splitter outputs.

Similar to the case for direct measurement considered in Section~S1,
the optical field at the start of the protocol is given by
\begin{equation}
\left | \psi_{\rm in} \right \rangle = \left (\sqrt{1-\xi} + \sqrt{\xi} \hat{e}^\dagger \right ) \hat{D}_{h}(\alpha_h) |0, \dots, 0\rangle,
\end{equation}
where $ |0, \dots, 0\rangle$ is shorthand for $| 0 \rangle_e | 0 \rangle_h | 0 \rangle_{v_1} | 0 \rangle_{v_2} | 0 \rangle_{d_1} | 0 \rangle_{d_2} | 0 \rangle_{l_1} | 0 \rangle_{l_2}$.
$\hat{D}_{h}(\alpha_h) = \exp (\alpha_h \hat{h}^\dagger - \alpha_h^* \hat{h})$ is a displacement operator that converts the field associated with $\hat{h}$ into a coherent state with amplitude $\alpha_h = \sqrt{\epsilon} \alpha_c$ (here, $\alpha_c$ is the coherent amplitude of the input coherent field). 

Following our approach for direct measurement in Section~1,
we substitute in for $\hat h$ and $\hat e$ in terms of the input annihilation operators using Eqs.~(\ref{HOM_e})~and~(\ref{HOM_h}). This yields the output state
\begin{equation}
\begin{split}
\left | \psi_{\rm out} \right \rangle = \left (\sqrt{1-\xi} + \sqrt{\xi} \left ( \sqrt{\frac{\eta}{2}} \left (\hat{d}_1^\dagger + \hat{d}_2^\dagger \right ) + \sqrt{\frac{1-\eta}{2}} \left ( \hat{l}_1^\dagger + \hat{l}_2^\dagger \right )  \right ) \right ) \times \label{psi_out_HOM} \\
\hat{D}_{d_1}(-\alpha_d) \hat{D}_{d_2}(\alpha_{d})  \hat{D}_{l_1}(-\alpha_{l})  \hat{D}_{l_2}(\alpha_{l})  |0, \dots, 0\rangle,
\end{split}
\end{equation}
where $\alpha_d=\sqrt{\eta \epsilon/2} \alpha_c$ and $\alpha_l=\sqrt{(1-\eta)\epsilon/2}  \alpha_c$.

Operating on vacuum with the displacement operator gives the coherent state
\begin{equation}
\hat{D} (\alpha) |0\rangle = | \alpha \rangle = e^{-{\bar{n}}/{2}} \sum_{m=0}^\infty \frac{\alpha^m}{\sqrt{m!}} | m \rangle,
\end{equation}
 where the mean photon number $\bar n = |\alpha|^2$. Using this in Eq.~(\ref{psi_out_HOM}) together with the ladder operator relations
 \begin{eqnarray}
\hat{a}^\dagger | k \rangle &=& \sqrt{k+1} | k+1 \rangle\\
 \sum_{k=0}^\infty \frac{\alpha^k}{\sqrt{k!}} \sqrt{k+1}  | k+1 \rangle &=&  \sum_{k=0}^\infty \frac{\alpha^{k-1}}{\sqrt{k!}} k  | k \rangle,
 \end{eqnarray}
  we obtain
 \begin{equation}
\begin{split}
\left | \psi_{\rm out} \right \rangle = e^{-\epsilon \bar{n}_c/2} \sum_{k,l,m,n=0}^\infty \frac{\alpha_{d}^{k+l} \alpha_{l}^{m+n}}{\sqrt{k! \, l! \, m! \, n!}}  \left [ \sqrt{1-\xi} + \sqrt{\frac{\xi}{\epsilon}}\left ( \frac{l-k + n-m}{\alpha_c}\right )\right ] | k \, l \,  m \, n \rangle
\end{split}
\end{equation}
where $| k \, l \,  m \, n \rangle \equiv  | k \rangle_{d_1}| l \rangle_{d_2}  | m \rangle_{l_1}|  n \rangle_{l_2}$. For succinctness we have neglected the input mode labels in the ket, since all of these are vacuum after the fields propagate through the system. Note that here, while the subscript $l$ on $\alpha_l$ labels that this is the coherent amplitude reaching the output loss modes, while the non-subscript $l$'s are indices for the sum.

The probability of simultaneously detecting $\{ k, l, m,n \}$ photons in the output modes $\{ d_1, d_2, l_1 , l_2 \}$ is then
\begin{eqnarray}
P_{klmn} &=& \left | \left \langle k l m n \big | \psi_{\rm out} \right \rangle \right |^2\\
&=& e^{-\bar{n}_h}  \frac{\bar{n}_{d}^{k+l} \bar{n}_{l}^{m+n}}{k! \, l! \, m! \, n!} \Bigg \{ 
 \left ( \sqrt{1-\xi} + \sqrt{\frac{\xi}{\epsilon}} \frac{(l-k)}{\alpha_c} \right )^2 + 2\sqrt{\frac{\xi}{\epsilon}} \Big ( \sqrt{1-\xi}  \nonumber \\
 && \hspace{30mm} + \sqrt{\frac{\xi}{\epsilon}} \frac{(l-k)}{\alpha_c} \Big )   \frac{(n-m)}{\alpha_c} + \frac{\xi}{\epsilon} \frac{(n-m)^2}{\bar{n}_c}  
\Bigg \},
\end{eqnarray}
where $\bar{n}_d = |\alpha_d|^2$ and  $\bar{n}_l = |\alpha_l|^2$. 

Tracing over the loss channels since they are inaccessible and re-expressing the mean photon numbers in terms of the input mean photon number $\bar{n}_c$, we obtain the probability of detecting $\{ k, l \}$ photons in the detector modes $\{ d_1, d_2 \}$
\begin{eqnarray}
\bar{P}_{kl}  &=&  \sum_{m,n = 0}^\infty P_{klmn} \\
 &=& \frac{e^{-\eta \epsilon \bar{n}_c}}{k!  l!} \left ( \frac{\eta \epsilon \bar{n}_{c}}{2} \right )^{k+l} \left \{ 
1- \eta \xi + \frac{\xi}{\epsilon} \frac{(l-k)^2}{\bar{n}_c} + 2 \cos \theta \sqrt{\frac{(1-\xi)\xi}{\epsilon \bar{n}_c}}(l-k) 
\right \}, \label{noise_free_dist}
\end{eqnarray}
where we have defined $\alpha_c = \sqrt{\bar{n}_c} \, e^{i \theta}$ and have made use of the relations
\begin{eqnarray}
\sum_{m=0}^\infty \frac{x^m}{m!} &=& \exp(x)\\
\sum_{m=0}^\infty \frac{m \, x^{2m}}{m!} &=& \exp(x^2) \, x^2\\
\sum_{m=0}^\infty \frac{m^2 x^{2m}}{m!} &=& \exp(x^2) \, x^2 (1+x^2).
\end{eqnarray}

The photon count probability distribution in Eq.~(\ref{noise_free_dist}) is accurate in the ideal limit that there are no noise counts. However, noise counts are introduced from three sources: the coherent light that is not overlapped with the emitter field but still arrives at each detector, background photons produced at the emitter, and counts due to electronic noise. The former of these has Poissionian statistics, and it is reasonable to assume the other two do too. They are independent noise channels, so that -- under the assumption that they are Poissionian, their variances add, with the total variance at each detector equal to the total mean noise photon counts at that detector $\bar{n}_n/2 = \eta (1-\epsilon) \bar{n}_c/2 + \eta  \bar{n}_e/2 + \bar{n}_i$. The probability of detecting $k$ noise photon counts at either detector is then described by the Poissionian probability distribution 
\begin{equation}
P_{k, \rm noise} = \frac{e^{-\bar{n}_n/2}}{k!} \left ( \frac{\bar{n}_n}{2} \right )^k.
\end{equation}
The probability distribution of photon counts in the presence of this Poissionian noise is found, in a similar way as for direct measurement, by convolving the noise free distribution of Eq.~(\ref{noise_free_dist}) with this noise distribution:
\begin{equation}
\bar{P}_{jk}^{\rm coh} = \sum_{p=0}^j \sum_{q=0}^k \bar{P}_{pq}  P_{j-p, \rm noise} P_{k-q, \rm noise}. 
\end{equation}
Performing the convolution, we find
\begin{equation}
\begin{split}
{p}_{jk}(\xi) = \frac{e^{-\bar{n}}}{j! k!} \left (\frac{\bar{n}}{2} \right )^{j+k} \Bigg \{ 1 - \eta \xi \Bigg [1 -  \frac{\bar{n}_n}{\bar{n}} \left (\frac{j+k}{\bar{n}}  \right )  - \eta \epsilon \, \bar{n}_c \left ( \frac{j-k}{\bar n} \right )^2 \\
 + 2 \cos \theta \sqrt{\left ( \frac{1-\xi}{\xi} \right ) \epsilon \, \bar{n}_c} \left ( \frac{j-k}{\bar{n}}\right )  \Bigg ]  \Bigg \}, \label{P_coh_supp}
 \end{split}
\end{equation}
where $\bar{n} =  \eta(\bar{n}_c + \bar{n}_e) + 2 \bar{n}_i$ is the total mean photon number detected over both detectors, excluding photons from the emitter. This includes clicks from electronic noise, $\bar{n}_c$ is the photon number in the coherent field, $\epsilon$ is the mode matching efficiency between coherent field and emitter field, and $\theta$ is the phase difference between the two fields. 

Several features of Eq.~(\ref{P_coh_supp}) are notable. First and perhaps most interesting, if the emitter emits a photon with 100\% probability and the detection is perfect ($\eta=\epsilon=\xi=1$, $\bar{n}=\bar{n}_c$), then $\bar{P}_{jj}^{\rm coh} =0$ -- i.e., {\it perfect HOM interferences between a single photon and a coherent field can never produce the same number of photons on each side of the beam splitter for any brightness of coherent field}. Second, an interference term is present between the emitted and coherent fields. This term provides the strongest information about the presence of the emitter. Its  magnitude is maximised when $|\cos \theta|=1$.
Third, if the emitter emits with unit probability ($\xi=1$), the interference term vanishes. This can be understood since, with perfect emission, there is no vacuum term in the state of the emission ($| e \rangle = \sqrt{1-\xi}|0\rangle + \sqrt{\xi} |1 \rangle$). With only a single term, the emission no longer has a well defined phase. Fourth, the interference term also vanishes when $k=j$. 

In the case of incoherent HOM, the phase $\theta$ is randomised. Averaging over this, we find the total photon number distribution for incoherent HOM
\begin{equation}
{p}_{jk}(\xi) = \frac{e^{-\bar{n}}}{j! k!} \left (\frac{\bar{n}}{2} \right )^{\! j+k} \left \{ 1 - \eta \xi \left [1 -  \frac{\bar{n}_n}{\bar{n}} \left (\frac{j+k}{\bar{n}}  \right )  - \eta \epsilon \, \bar{n}_c \left ( \frac{j-k}{\bar n} \right )^2  \right ]  \right \}. 
\label{P_incoh}
\end{equation}

\subsection{b. Bayesian estimation}

Now that we have obtained the photon statistics for both incoherent and coherent HOM, we can follow the same approach as for direct measurement to determine the likelihood ratio for each possible measurement outcome and perform Bayesian inference, thereby deciding whether the emitter is present for a given sequence of measurements. 

Using Eq.~(\ref{lambda_j_eqn}), we find
\begin{eqnarray}
\lambda_{jk}^\text{coh} &=&   \frac{\bar{P}_{jk}^\text{coh}(0) }{\bar{P}_{jk}^\text{coh}(\xi) } \\
&=&   \left \{ 1 - \eta \xi  \left [1 -  \frac{\bar{n}_n}{\bar{n}}  \left (  \frac{j+k}{\bar{n}}   \right )  - \eta \epsilon  \bar{n}_c  \left (  \frac{j-k}{\bar n}  \right )^2   
 + 2 \cos \theta \sqrt{ \left ( \frac{1-\xi}{\xi}  \right ) \epsilon \bar{n}_c}  \left (  \frac{j-k}{\bar{n}}  \right )    \right ]   \right \}^{-1}.
\end{eqnarray}
The mean and standard deviation of the log-likelihood ratio can then be determined using Eqs.~(\ref{mu_x_e})~and~(\ref{sigma_x_e}), but now summing over both $j$ and $k$. Then the number of measurements required to reach a certain level of confidence can be determined following the approach in Section~S2{\bf b}.

\section*{S4. Optimal coherent photon number for emitter detection using coherent HOM with saturating detectors}

Figure 3{\bf c} in the main text shows the speed-up factor as a function of detection efficiency ($\eta$) and background noise $\hat{n}_e$ for saturating detectors, saturating at each of four, two and one photons. The speed-up factor is optimised over the coherent photon number $\bar{n}_c$ at each $\eta$ and $\hat{n}_e$. Figure~\ref{fig:nc_opt} in this supplement shows this optimal coherent photon number $\bar{n}_c$ for each of the three saturation lelels.

\begin{figure}[h!]
	\begin{center}
	\includegraphics[scale=0.5]{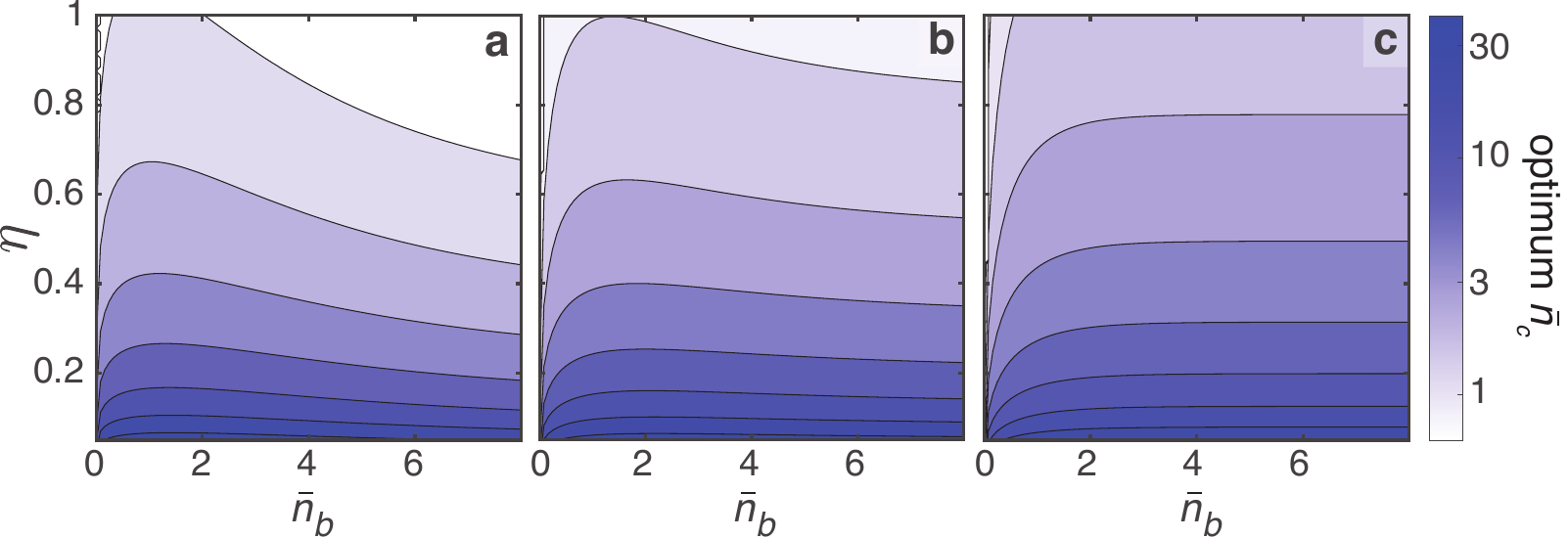}
	\caption{Optimal coherent photon number for coherent HOM with saturating photon number resolving detectors as a function of $\bar{n}_e$ and $\eta$, with $\epsilon=0.9$ and $\xi=0.1$. {\bf a.} Saturating at $t=4$ photons. {\bf b.} Saturating at $t=2$ photons.  {\bf c.} Saturating at $t=1$ photon. }
	\label{fig:nc_opt}%
	\end{center}
\end{figure}

\end{document}